\documentclass[aps,prd,onecolumn,eqsecnum,amsmath,nofootinbib,preprintnumbers,superscriptaddress]{revtex4-2}

\usepackage[dvipsnames]{xcolor}
\usepackage{color,graphicx,float,subfigure,xcolor}
\usepackage{amsfonts,amssymb,mathrsfs,times}
\usepackage{bm}
\usepackage{multirow}
\usepackage{mathtools}
\usepackage{dsfont}
\usepackage{setspace}
\usepackage{amsmath} 
\usepackage{graphicx}
\usepackage[colorlinks,linkcolor=blue,anchorcolor=blue,citecolor=blue]{hyperref}
\usepackage{ulem}
\usepackage{subcaption}
\usepackage{amsthm}

\usepackage{subcaption}
\usepackage[justification=raggedright,singlelinecheck=false]{caption}

\begin{document}
\thispagestyle{empty}
% \preprint{\hfill {\small {ICTS-USTC/PCFT-24-03}}}
%<<<<<<<<<<<<< TITLE >>>>>>>>>>>>>>>%
\title{The (in)stability on total transmission modes with small bumps}
	
%<<<<<<<<<<<<< AUTHOR >>>>>>>>>>>>>>>%
%\author{$^b$}
%
%\email{}

\author{Xi-Ning Dong}
\email{dongxining24@mails.ucas.ac.cn}
\affiliation{School of Fundamental Physics and Mathematical Sciences, Hangzhou Institute for Advanced Study, UCAS, Hangzhou 310024, China}
\affiliation{CAS Key Laboratory of Theoretical Physics, Institute of Theoretical Physics, Chinese Academy of Sciences, Beijing 100190, China}
\affiliation{University of Chinese Academy of Sciences, Beijing 100190, China}

\author{Liang-Bi Wu}
\email[corresponding author: ]{wulb@ucas.ac.cn}
\affiliation{School of Fundamental Physics and Mathematical Sciences, Hangzhou Institute for Advanced Study, UCAS, Hangzhou 310024, China}

\author{Yu-Sen Zhou}
\email{zhou\_ys@mail.ustc.edu.cn}
\affiliation{Interdisciplinary Center for Theoretical Study and Department of Modern Physics,\\
	University of Science and Technology of China, Hefei, Anhui 230026, China}

\author{Zong-Kuan Guo}
\email{guozk@itp.ac.cn}
\affiliation{CAS Key Laboratory of Theoretical Physics, Institute of Theoretical Physics, Chinese Academy of Sciences, Beijing 100190, China}
\affiliation{University of Chinese Academy of Sciences, Beijing 100190, China}
\affiliation{School of Fundamental Physics and Mathematical Sciences, Hangzhou Institute for Advanced Study, UCAS, Hangzhou 310024, China}

%<<<<<<<<<<<<< DATE >>>>>>>>>>>>>>>%
\date{\today}
	
%======================================%
%<<<<<<<<<<<<< ABSTRACT >>>>>>>>>>>>>>>%
%======================================%
\begin{abstract}
Total transmission modes (TTMs) are a class of reflectionless solutions to black hole perturbation equations, closely related to quasinormal modes (QNMs), and can exhibit significant sensitivity to weak environmental perturbations. In this work, we investigate the spectrum (in)stability of TTMs of Tangherlini black holes by introducing a localized P\"{o}schl-Teller bump perturbation into the effective potential, and employ a modified Chebyshev-Lobatto grid to improve the numerical accuracy of the localized perturbation. For $d=14$, $\ell=2$, and $s=2$, the purely imaginary TTM exhibits relatively strong spectrum stability, whereas the genuine complex TTMs undergo significant migrations even for small perturbations, consistent with the spectrum stability revealed by previous pseudospectrum analyses.
\end{abstract}

\maketitle
	
%======================================%
%<<<<<<<<<<< Introduction >>>>>>>>>>>>>%
%======================================%	
\section{Introduction}\label{Introduction}
Since the first direct detection of gravitational waves (GWs) by LIGO in 2015~\cite{LIGOScientific:2016aoc,LIGOScientific:2020ibl}, gravitational-wave astronomy has become an indispensable tool for probing compact objects and testing gravitational theories, because gravitational waves carry abundant information about astrophysical systems and the underlying nature of gravity. For the coalescence of binary black holes, the gravitational-wave signal can be broadly divided into the ``inspiral'', ``merger'', and ``ringdown'' stages. Especially in the ringdown stage, the gravitational-wave signal is dominated by quasinormal modes (QNMs), whose frequencies and damping times depend solely on the intrinsic properties of the remnant black hole, such as its mass and spin. Therefore, the study of black hole QNMs provides a powerful tool for extracting the physical parameters of black holes, testing General Relativity (GR), and exploring possible new physics beyond Einstein's theory. This approach is commonly referred to as black hole spectroscopy~\cite{Kokkotas:1999bd,Berti:2009kk,Konoplya:2011qq,Berti:2025hly}.

From the point of view of scattering theory, QNMs, which correspond to poles of the reflection and transmission coefficients, are not the only class of modes. Another important family is known as total transmission modes (TTMs), for which the reflection coefficient vanishes, allowing an incident wave to be transmitted through the effective potential barrier without reflection. For an isolated black hole, the TTM spectrum is determined entirely by the intrinsic black hole parameters, making TTMs a potentially valuable complementary probe of black hole properties. Like QNMs, TTMs are complex-frequency solutions of the perturbation equation, but they satisfy different asymptotic boundary conditions. QNMs are purely ingoing at the event horizon and purely outgoing at spatial infinity, whereas TTMs are purely outgoing ($\text{TTM}_{\text{L}}$) or purely ingoing ($\text{TTM}_{\text{R}}$) at both asymptotic boundaries. TTMs have been extensively investigated in both static and rotating black hole spacetimes through analytical and numerical approaches~\cite{Couch:1973zc,Wald:1973wwa,Chandrasekhar:1984mgh,Andersson:1994tt,MaassenvandenBrink:2000iwh,Chandrasekhar:1984mgh,Berti:2004md,Keshet:2007be,Cook:2016fge,Cook:2016ngj,Cook:2018ses,Cook:2022kbb}. More recently, TTMs are also investigated in various spacetimes~\cite{Qian:2025occ,Chen:2025sbz} and  draining bathtub model~\cite{Yu:2026rku}. The TTM can be selectively excited by an incident wave designed to grow exponentially over time, with its growth rate precisely determined by the imaginary part of the mode. In this situation, the reflected wave does not emerge until the exponential amplification is unavoidably stopped by computational limitations or experimental boundaries~\cite{Tuncer:2025dnp,Wu:2026hvf}. This phenomenon is known as virtual absorption, which provides a distinctive physical manifestation of TTMs in black hole scattering. Together with the dependence of their characteristic frequencies on the underlying black hole spacetime, this remarkable scattering behavior makes TTMs potentially valuable probes of spacetime geometry.

The open boundary conditions of QNMs render the corresponding spectrum problem intrinsically non-Hermitian (NH)~\cite{Ashida:2020dkc}, with spectrum instability constituting an important issue in the study of NH gravitational systems. In recent years, pseudospectrum analysis has been widely employed to characterize the instability of QNM spectra (see e.g. \cite{Jaramillo:2020tuu,Destounis:2021lum,Cao:2024oud,Chen:2024mon,Cao:2024sot,Cao:2025qws,Cai:2025irl} and references therein). Since TTMs possess a similar open NH structure, their spectrum (in)stability can likewise be investigated within the pseudospectrum framework. Recent work \cite{Zhou:2025xdo} has extended pseudospectrum analysis to TTMs of Tangherlini black black holes. Beyond pseudospectrum analysis, spectrum (in)stability can also be directly probed by introducing small perturbations to the effective potential and tracking the resulting migration of the spectrum~\cite{Cheung:2021bol,Xie:2025jbr,Yang:2024vor,Courty:2023rxk,Cardoso:2024mrw,Ianniccari:2024ysv,Hu:2025efp,MalatoCorrea:2025iuc}. It is known that in a four-dimensional Schwarzschild black hole, for a fixed value of $\ell$, there are only two purely imaginary TTMs. As a result, the four-dimensional case is less suitable for analyses of spectrum (in)stability. The Tangherlini black hole, a straightforward higher-dimensional generalization of the Schwarzschild black hole, possesses genuinely complex TTMs~\cite{Tuncer:2025dnp,Zhou:2025xdo}. In this paper, we adopt this approach to investigate the spectrum (in)stability of TTMs in the Tangherlini black hole, considering generic small perturbations to the effective potential that may arise from matter in the local black hole environment~\cite{Barausse:2014tra}.

The paper is organized as follows. In Sec. \ref{Set up}, we first review the perturbation equations for Tangherlini black holes and define TTMs according to their asymptotic behavior. We also introduce small perturbation of the effective potential consisting of tiny bumps, which may be produced by matter in the local black hole environment. In Sec. \ref{Result}, we show the numerical results of the migrations of TTM spectra. Sec. \ref{sec: conclusions} is the conclusions and discussion. In Appendix \ref{app: methods}, we introduce the Eddington-Finkelstein coordinates and compactified radial coordinates to recast the TTM problem into a generalized eigenvalue problem. Appendix \ref{app: modified CL} presents the modified Chebyshev-Lobatto grid.

\section{Set up}\label{Set up}
We consider the Tangherlini black hole, the higher-dimensional generalization of the Schwarzschild black hole in $d$ dimensions~\cite{Tangherlini:1963bw}. Its metric can be written as
\begin{eqnarray}\label{metric}
    \mathrm{d}s^{2}=-f(r)\mathrm{d}t^{2}+f(r)^{-1}\mathrm{d}r^{2}+r^{2}\mathrm{d}\Omega_{d-2}^{2}\, ,
\end{eqnarray}
where $\mathrm{d}\Omega_{d-2}^{2}$ is the line element on the $(d-2)$-dimensional unit sphere,
\begin{eqnarray}\label{metric_function}
  f(r)=1-\Big(\frac{r_{\text{h}}}{r}\Big)^{d-3}\, ,
\end{eqnarray}
and $r=r_{\text{h}}$ is the location of the event horizon. The perturbation of this black hole is described by the following master equation~\cite{Kodama:2003jz, Kodama:2003kk}
\begin{eqnarray}\label{masterEqTime}
    \Big[\frac{\partial^2}{\partial t^2}-\frac{\partial^2}{\partial x^2}+V(r)\Big]\Psi(t,x)=0\, ,
\end{eqnarray}
where the tortoise coordinate $x$ is defined as $\mathrm{d}x=f(r)^{-1}\mathrm{d}r$ and the effective potential,
\begin{eqnarray}\label{potential}
    V(r)=\frac{f(r)}{4r^2}\Bigg[4\ell(\ell+d-3)+(d-2)(d-4)+(1-s^2)(d-2)^2\Big(\frac{r_\text{h}}{r}\Big)^{d-3}\Bigg]\, ,
\end{eqnarray}
depends on the angular multipole number $\ell$, with $s=0$ describing either massless scalar or gravitational tensor perturbations, $s=2$ corresponds to gravitational vector perturbations, and $s=2/(d-2)$ and $s=2-2/(d-2)$ to the electromagnetic vector and scalar perturbations, respectively. To work in the frequency domain, we introduce a Fourier decomposition $\Psi(t,x)=\mathrm{e}^{-\mathrm{i}\omega t}\psi(x)$ in Eq. (\ref{masterEqTime}),
\begin{eqnarray}\label{masterEq}
    \Big[\frac{\mathrm{d}^2}{\mathrm{d} x^2}+\omega^2-V(r)\Big]\psi=0\, ,
\end{eqnarray}
TTMs are defined by their asymptotic behaviors at two boundaries. The right TTM ($\text{TTM}_\text{R}$) and left TTM ($\text{TTM}_\text{L}$) are defined as~\cite{Tuncer:2025dnp,Zhou:2025xdo}:
\begin{eqnarray}\label{boundary_conditions}
    \text{TTM}_{\text{R}}: \quad {\psi} &\sim \mathrm{e}^{-\mathrm{i}\omega x},
    \qquad \text{as} \quad x\to \pm\infty\, ,\nonumber\\
    \text{TTM}_{\text{L}}: \quad {\psi} &\sim \mathrm{e}^{+\mathrm{i}\omega x},
    \qquad \text{as} \quad x\to \pm\infty\, .
\end{eqnarray}

In Ref. \cite{Zhou:2025xdo}, the (in)stability of TTMs has been studied by the pseudospectrum. However, the information provided by pseudospectra regarding spectrum stability is insufficient. This can be understood from its definition: for a given perturbation magnitude under a specific norm, the maximal spectrum shift precisely corresponds to the boundary of the pseudospectrum. Although it is valuable, its application is greatly limited by the fact that it cannot pinpoint the exact locations of spectrum shifts. Therefore, like modifying the effective potential to study the QNM spectrum instability~\cite{Cheung:2021bol,Xie:2025jbr,Yang:2024vor,Courty:2023rxk,Cardoso:2024mrw,Ianniccari:2024ysv,Hu:2025efp,MalatoCorrea:2025iuc}, we modify the effective potential (\ref{potential}) to study the (in)stability of TTMs. 

Consider a small perturbation to the effective potential of the form
\begin{eqnarray}\label{s perturbation}
    V_\epsilon(x)=V(x)+\epsilon V_{\text{bump}}(x)\, ,
\end{eqnarray}
where $\epsilon \ll 1$ and $V_{\text{bump}}(x)$ stands for a generic bump which decays at least faster than $V(x)$ as $x\to \pm\infty$. In this work, we focus on the case with $V_{\text{bump}}$ being the P\"{o}schl-Teller (PT) form in terms of tortoise coordinate $x$, namely 
\begin{eqnarray}\label{pre:PT}
    V_\text{bump}(x)=\text{sech}^2\Big[\alpha(x-a)\Big]\, ,
\end{eqnarray}
in which $\alpha>0$ indicates width of the bump, and $a$ is the position of the bump. The larger $\alpha$, the narrower the bump, while the smaller $\alpha$, the wider the bump. Using the ingoing and outgoing Eddington-Finkelstein (EF) coordinate, TTM problem can be transformed into a general eigenvalue problem. One can refer to more detail on this in Appendix \ref{app: methods}. More importantly, if we discretize the perturbed operator using the standard Chebyshev-Lobatto grid, the perturbed TTM spectrum is very difficult to be computed accurately. For this type of model with a small bump, we adopt a modified Chebyshev-Lobatto (CL) grids to discretize the operator (see Appendix \ref{app: modified CL}). In doing so, the numerical results converge much better.

\begin{figure}[htbp]
    \centering
    \includegraphics[width=0.69\textwidth]{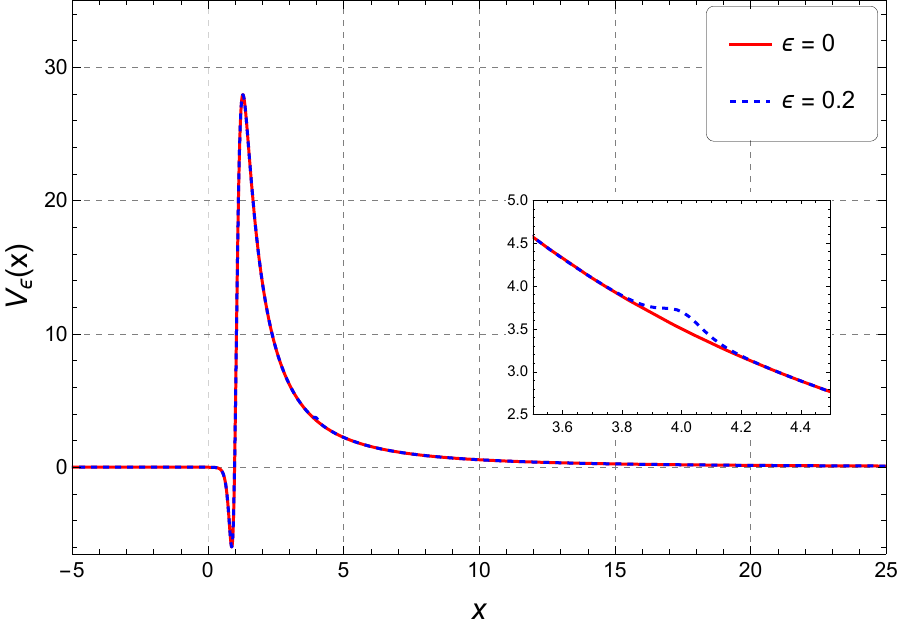}
    \caption{The comparison‌ of the original potential and the modified potential with a P\"oschl-Teller bump, where red solid line stands for original potential and blue dashed line stands for modified potential with $\epsilon=0.2$.}
    \label{fig:potential}
\end{figure}

It should be noted that not all $\alpha>0$ can meet the requirements of physics. At the end of this section, we compare the asymptotic behaviors of $V$ and $V_\text{bump}$ to further constrain $\alpha$. Near the event horizon $x\to-\infty$, the tortoise coordinate behave as
\begin{eqnarray}
    x\sim\frac{r_\text{h}}{d-3}\ln\Big(\frac{r-r_\text{h}}{r_\text{h}}\Big)\, ,\qquad  r\to r_\text{h}^+\, ,
\end{eqnarray}
and therefore the bump in Eq. (\ref{pre:PT}) behaves as 
\begin{eqnarray}
    V_{\text{bump}}\sim(r-r_\text{h})^{\frac{2\alpha r_\text{h}}{d-3}}\, ,
\end{eqnarray}
while the original effective potential behaves as $V(x)\sim (r-r_{\mathrm{h}})$. Thus, the requirement of $V_{\text{bump}}$ decaying at least as fast as $V$ at means that
\begin{eqnarray}\label{condition_alpha}
    \alpha r_\text{h}\geq\frac{d-3}{2}\, .
\end{eqnarray}
For the spatial infinity, such physical requirement satisfies automatically. In the followings, we always use the unit with $r_{\text{h}}=1$. In Fig. \ref{fig:potential}, we show $V(x)$ and $V_{\epsilon}(x)$ for the original potential parameters $d=14$, $\ell=2$, $s=2$ (then $\alpha\geq11/2$), and the bump parameters $\epsilon=0.2$, $\alpha=11$, $a=4$.

%%%%%%%%%%%%%%%%%%%%%%%%%%%%%%%%%%%%%%%%%%%%%%%%%%%%%%%%%%%%%%%%%%%%
\section{The (in)stability of total transmission modes}\label{Result}
%%%%%%%%%%%%%%%%%%%%%%%%%%%%%%%%%%%%%%%%%%%%%%%%%%%%%%%%%%%%%%%%%%%%

In this section, we investigate the spectrum (in)stability of TTMs by studying the migration of the TTM spectra with changing the position of the bump $V_{\text{bump}}$. Ref. \cite{Tuncer:2025dnp} obtained the TTMs of Tangherlini black holes for {$s=0$ and $s=2$, and showed that purely imaginary TTMs appear for $d\geq4$ and exist only for $s=2$. In contrast, complex TTMs with nonvanishing real parts appear for $d\geq10$ in both the $s=0$ and $s=2$ cases. Ref. \cite{Zhou:2025xdo} further showed that such complex TTMs can already appear for $d\geq8$, and investigated the spectrum (in)stability of TTMs in Tangherlini black holes using pseudospectrum and eigenvalue condition numbers. In particular, three TTM spectra for $d=14$, $\ell=2$, and $s=2$ were analyzed in detail. Following the labeling convention in~\cite{Zhou:2025xdo}, three modes are denoted by $\omega_n (n=0,1,2)$ in decreasing order of their imaginary parts. Among them, $\omega_0$ is purely imaginary and lies on the positive imaginary axis, whereas $\omega_1$ and $\omega_2$ lie in the lower half of the complex-frequency plane and have nonvanishing real parts. It should be noted that, unlike QNMs, for which a positive imaginary part indicates dynamics instability, TTMs are allowed to have spectrum with positive imaginary parts. In the followings, we focus on the $\text{TTM}_{\text{L}}$ spectra for $d=14$, $s=2$, and $\ell=2$, with particular attention to the continuous migration trajectories of $\omega_1$ and $\omega_2$ in the complex-frequency plane as $a$ increases.

Fig. \ref{fig:migration with alpha_11} shows the variation of the spectrum shift $|\Delta\omega_0|$ of the $n=0$ TTM with the bump location $a$, together with the migration trajectories of the $n=1$ and $n=2$ TTMs in the complex-frequency plane for $\alpha=11$. For each value of $\epsilon$ labeled in the figure, we trace the corresponding TTM migration trajectory as the bump is moved outward from $a=1.3$, near the maximum of the original effective potential, to $a=4$. At $a=1.3$, the modes lie close to the unperturbed TTM spectrum ($\epsilon=0$), shown by black points, while the remaining markers indicate selected intermediate bump locations.

\begin{figure}[htbp]
    \centering
    \subfigure[]{\includegraphics[width=0.47\textwidth]{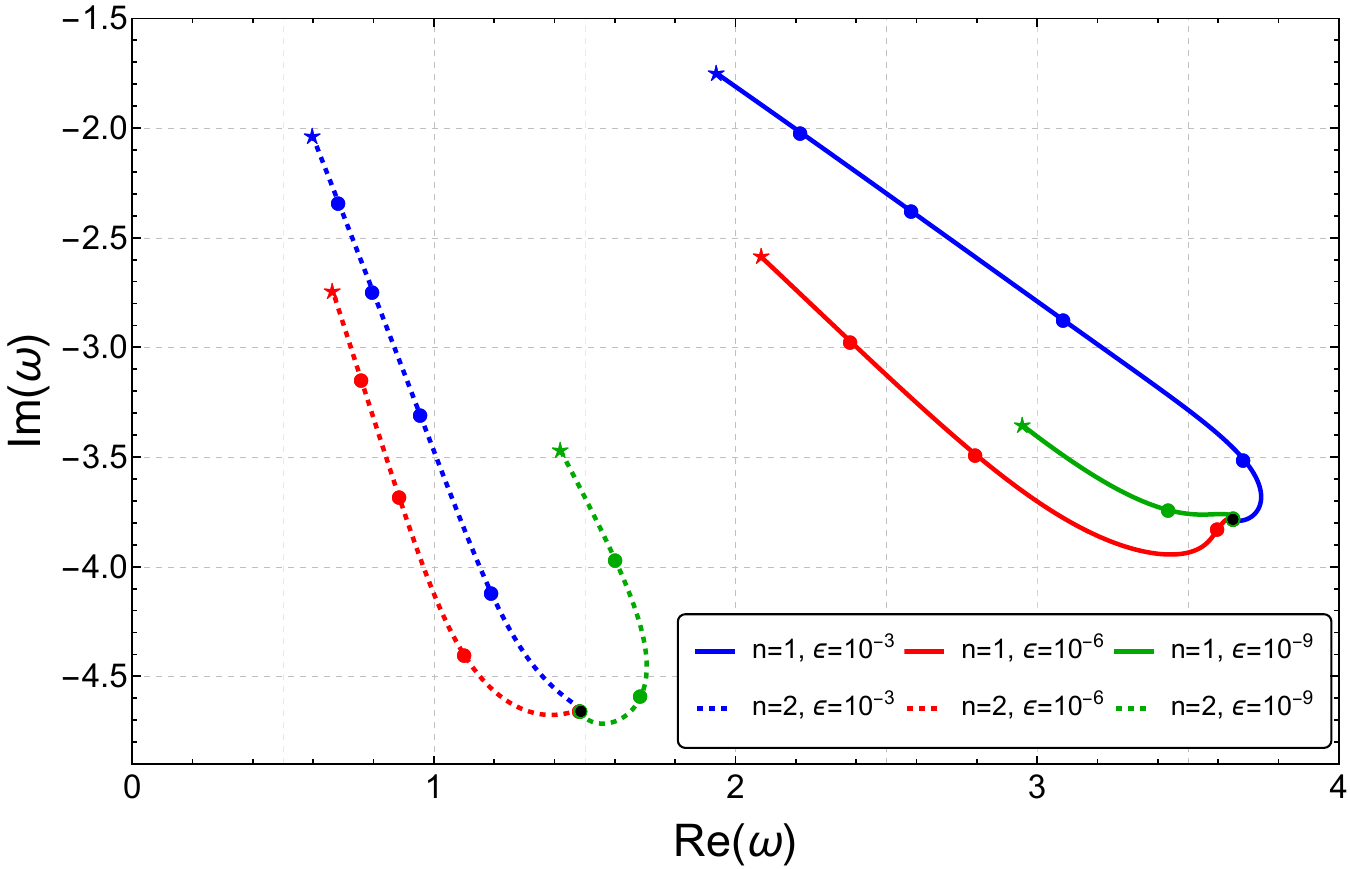}\label{fig:migration with alpha_11_1}}
    \hfill
    \subfigure[]{\includegraphics[width=0.47\textwidth]{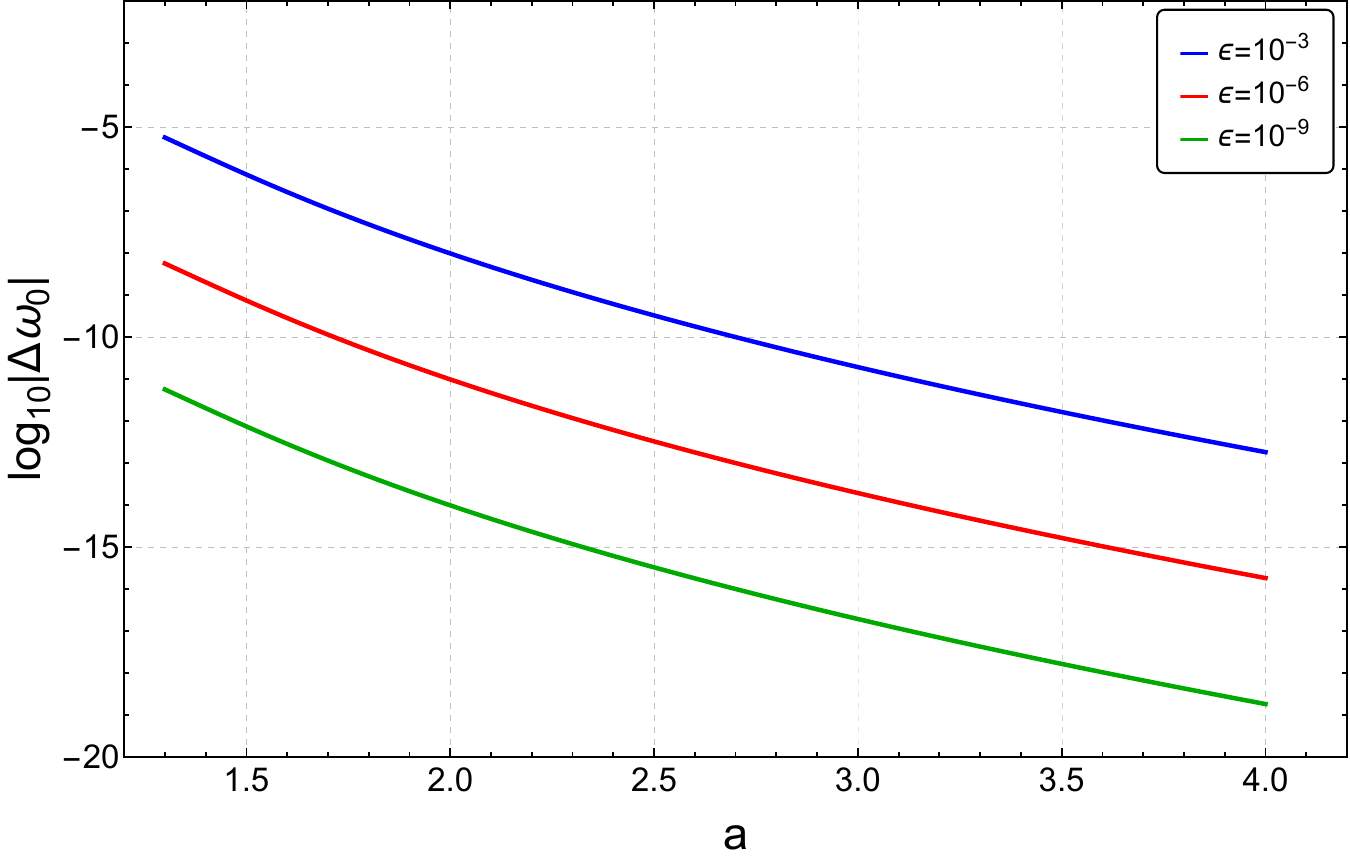}\label{fig:migration with alpha_11_2}}\\
    \subfigure[]{\includegraphics[width=0.47\textwidth]{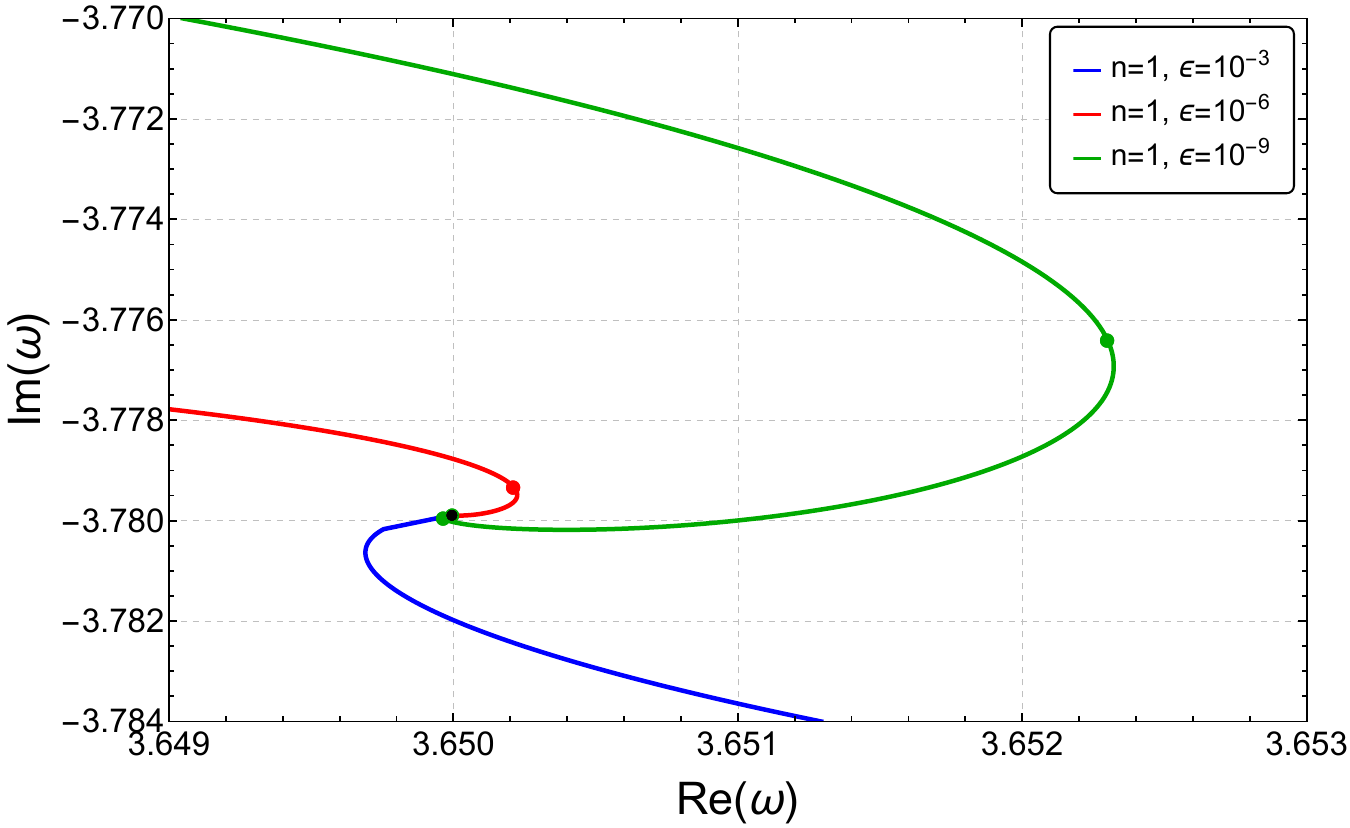}\label{fig:migration with alpha_11_3}}
    \hfill
    \subfigure[]{\includegraphics[width=0.47\textwidth]{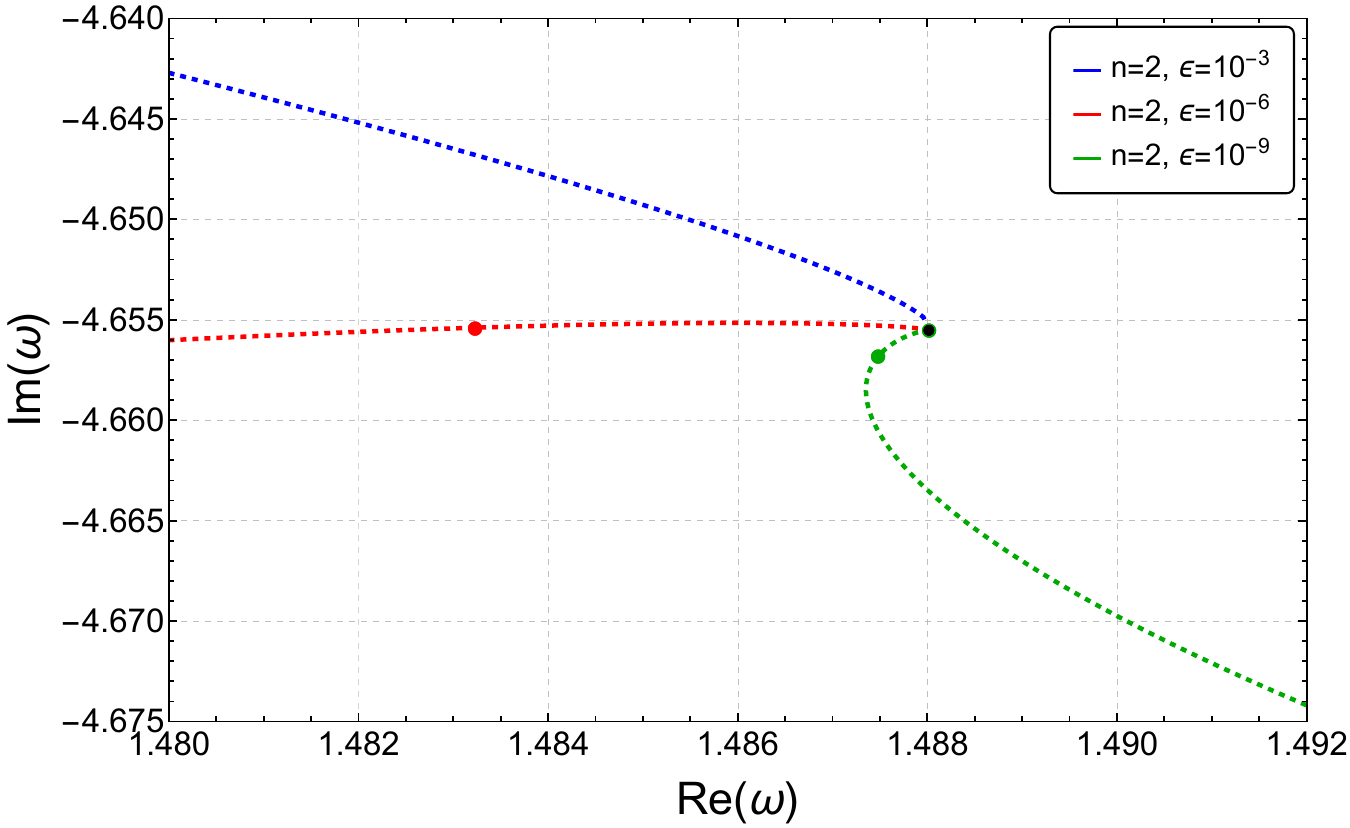}\label{fig:migration with alpha_11_4}}
\caption{Spectral migrations of the $n=0$, $n=1$, and $n=2$ TTMs induced by varying the bump location over $a\in[1.3,4]$ for $\alpha=11$, where black points denote the TTM spectra of original potential. (a) Migration trajectories of the $n=1$ and $n=2$ TTMs in the complex-frequency plane for different perturbation amplitudes $\epsilon$. The stars indicate the endpoints at $a=4$, while the colored points mark selected locations along the trajectories, namely $a=2$, $2.5$, $3$ and $3.5$. (b) Dependence of $\log_{10}|\Delta\omega_0|$ on the bump location $a$ for different values of $\epsilon$, where $\Delta\omega_0$ is defined as the difference between $\omega_0(a)$ and the original $\omega_0$. Panels (c) and (d) are the magnified views of the $n=1$ and $n=2$ trajectories in panel (a), respectively, showing the detailed spectral migration near the corresponding unperturbed TTM spectra. These numerical results are obtained from the resolution $N=400$, $\beta=8$ [see Eq. (\ref{q_x})], and a convergence criterion of $10^{-5}$ (Comparing the numerical results with $N=400$ and $N=450$).}

\label{fig:migration with alpha_11}
\end{figure}

Specifically, Fig. \ref{fig:migration with alpha_11_1} shows the migration trajectories of the $n=1$ and $n=2$ modes. At small $a$, the perturbed spectra remain in the vicinity of their corresponding unperturbed values. As the bump is gradually displaced outward, both modes depart continuously from the original spectral points and trace pronounced trajectories. The migration is clearly nontrivial: both the real and imaginary parts of the TTM spectra vary as $a$ increases, and the trajectories develop appreciable curvature rather than following a simple linear displacement. A particularly notable feature is that the migration remains clearly visible even when the perturbation amplitude is reduced to $\epsilon=10^{-9}$. Although the modification of the effective potential is extremely small, the corresponding $n=1$ and $n=2$ modes can still move appreciably away from their unperturbed values as $a$ increases. Moreover, changing $\epsilon$ modifies not only the overall extent of the migration but also the detailed trajectories followed by the modes. To resolve the initial stage of the migration, Figs. \ref{fig:migration with alpha_11_3} and \ref{fig:migration with alpha_11_4} provide enlarged views around the unperturbed $n=1$ and $n=2$ modes, respectively. The close-up views show how the perturbed branches emerge from the immediate neighborhood of the original TTMs as $a$ is increased. These two panels therefore complement the global trajectories by resolving the local deformation of the TTM spectrum near the unperturbed case.

As for the pure imaginary mode, TTM displays a qualitatively different response to $a$. Upon perturbation, such a mode remains on the imaginary axis, with an imaginary part larger than that of the corresponding unperturbed mode; its migration can therefore be conveniently characterized by $\log_{10}|\Delta\omega_0|$ as shown in Fig. \ref{fig:migration with alpha_11_2}. It can be seen that for a fixed $\epsilon$, the displacement decreases as the bump is moved outward. This is an important characteristic of the stability of this mode, which is consistent with the conclusions given in~\cite{Zhou:2025xdo}. In addition, reducing $\epsilon$ from $10^{-3}$ to $10^{-6}$ and subsequently to $10^{-9}$ produces an approximately corresponding downward shift of the curves.
\begin{figure}[htbp]
    \centering
    \subfigure[]{\includegraphics[width=0.47\textwidth]{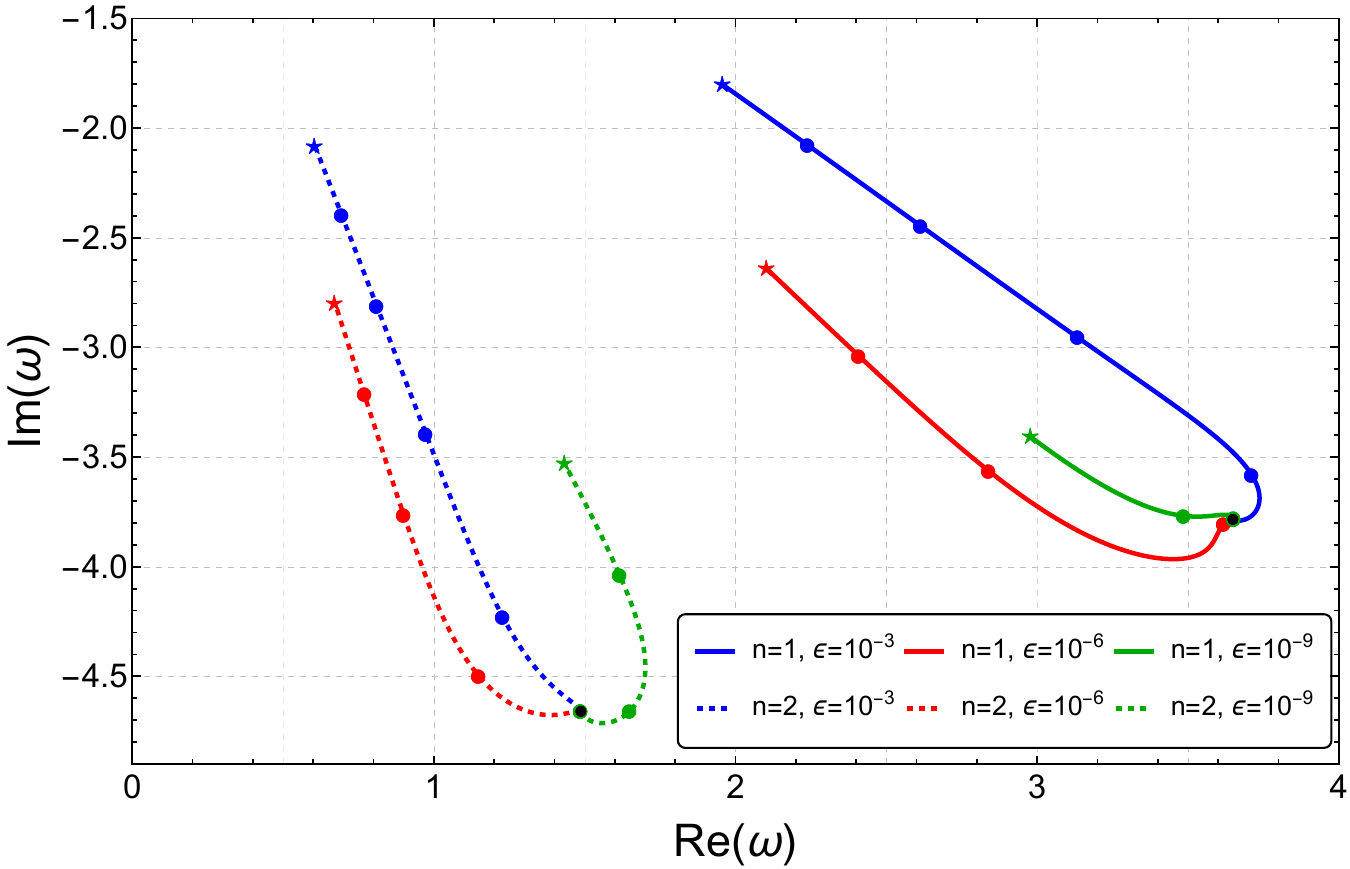}\label{fig:migration with alpha_16.5_1}}
    \hfill
    \subfigure[]{\includegraphics[width=0.47\textwidth]{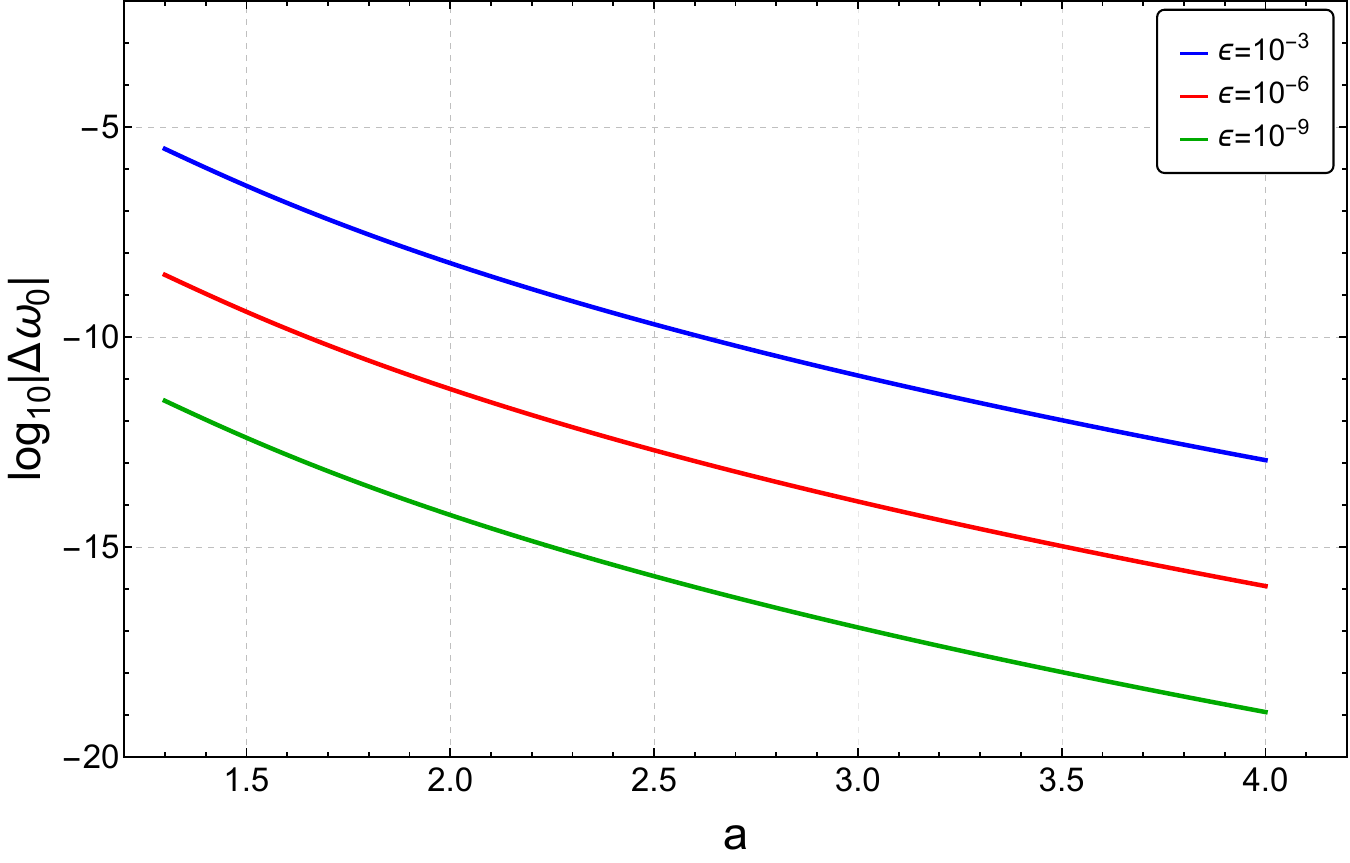}\label{fig:migration with alpha_16.5_2}}\\
    \subfigure[]{\includegraphics[width=0.47\textwidth]{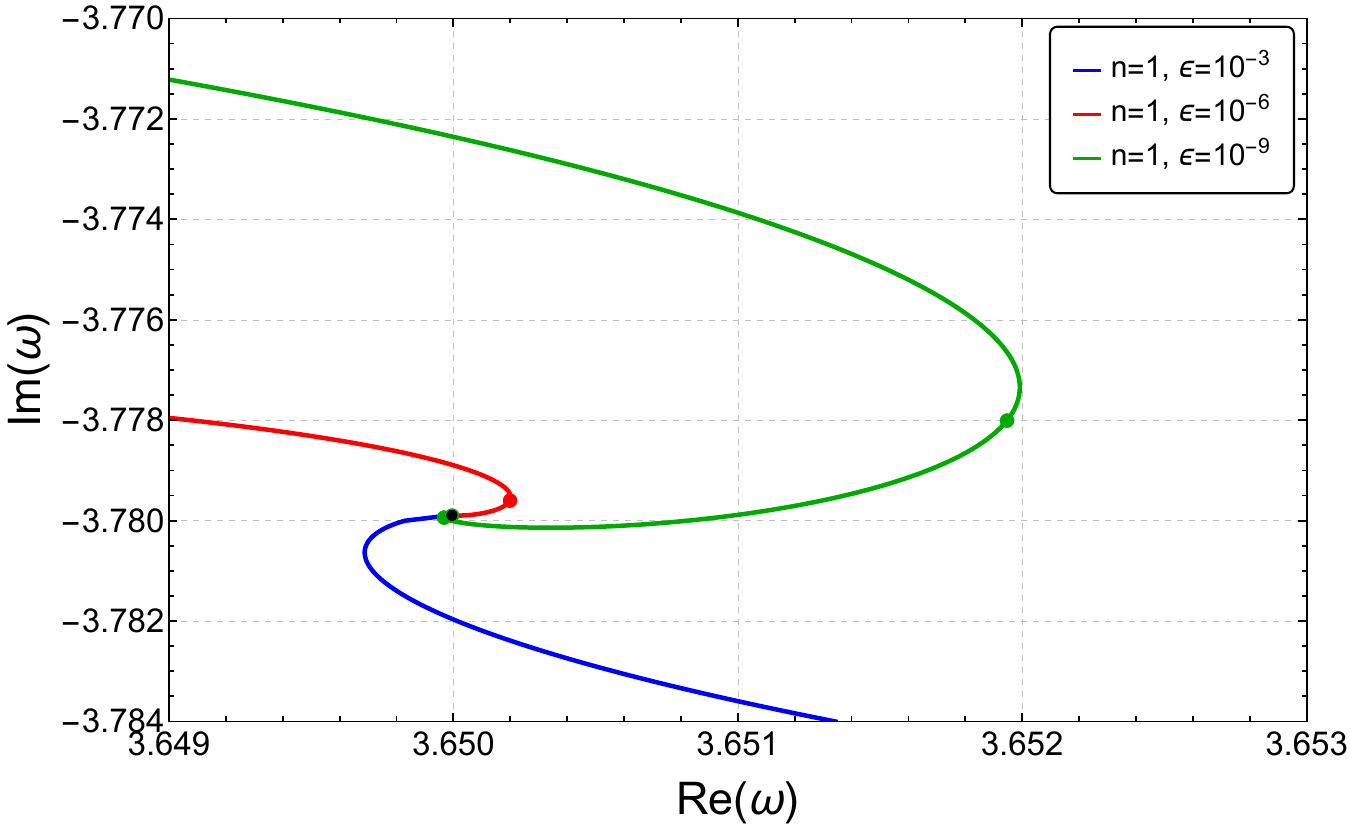}\label{fig:migration with alpha_16.5_3}}
    \hfill
    \subfigure[]{\includegraphics[width=0.47\textwidth]{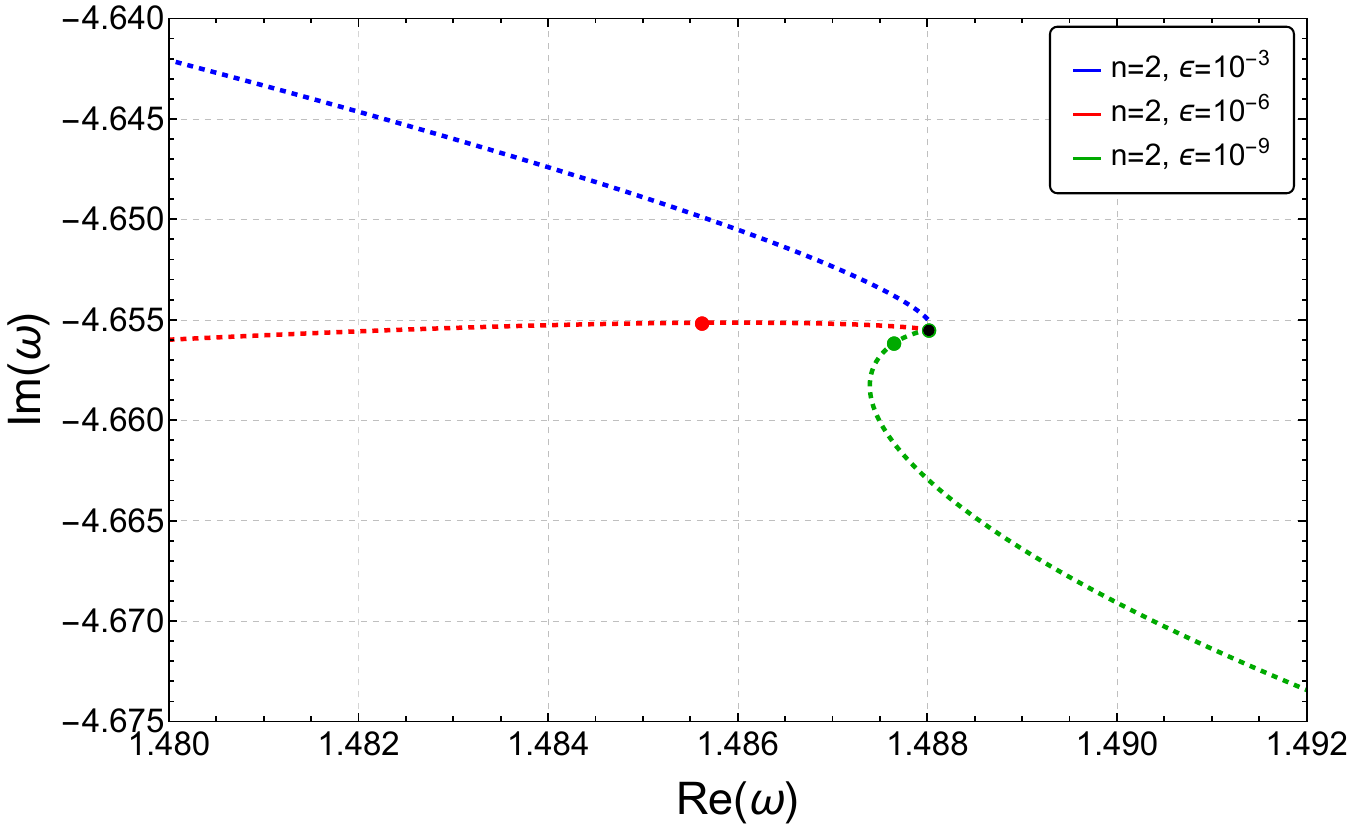}\label{fig:migration with alpha_16.5_4}}
\caption{Spectral migrations of the $n=0$, $n=1$, and $n=2$ TTMs induced by varying the bump location over $a\in[1.3,4]$ for $\alpha=16.5$, where black points denote the TTM spectra of original potential. (a) Migration trajectories of the $n=1$ and $n=2$ TTMs in the complex-frequency plane for different perturbation amplitudes $\epsilon$. The stars indicate the endpoints at $a=4$, while the colored points mark selected locations along the trajectories, namely $a=2$, $2.5$, $3$ and $3.5$. (b) Dependence of $\log_{10}|\Delta\omega_0|$ on the bump location $a$ for different values of $\epsilon$, where $\Delta\omega_0$ is defined as the difference between $\omega_0(a)$ and the original $\omega_0$. Panels (c) and (d) are the magnified views of the $n=1$ and $n=2$ trajectories in panel (a), respectively, showing the detailed spectral migration near the corresponding unperturbed TTM spectra. These numerical results are obtained from the resolution $N=500$, $\beta=9$ [see Eq. (\ref{q_x})], and a convergence criterion of $10^{-5}$ (Comparing the numerical results with $N=500$ and $N=550$).}
\label{fig:migration with alpha_16.5}
\end{figure}

In Fig. \ref{fig:migration with alpha_16.5}, as $\alpha$ is increased from $11$ to $16.5$, the overall shapes of the TTM spectral migration trajectories remain essentially unchanged for all three perturbation amplitudes, $\epsilon=10^{-3}$, $10^{-6}$, and $10^{-9}$, while the magnitude of the spectral migration decreases slightly. In contrast, variations in the bump location $a$ and perturbation amplitude $\epsilon$ lead to more pronounced frequency shifts. These results indicate that, within the parameter range considered here, the TTM spectral migration exhibits a relatively weak dependence on the width parameter $\alpha$.

TTMs studied here exhibit spectrum sensitivity similar to that of QNMs~\cite{Cheung:2021bol,Jia:2026vnx} under localized perturbations of the effective potential: even a small perturbation can induce a significant frequency shift as it is moved outward. However, their migration trajectories differ markedly. QNMs can develop characteristic spiral structures, whereas no analogous behavior is observed for the TTMs considered here.

%======================================%
%<<<<<<<<<<< Conclusions >>>>>>>>>>>>>%
%======================================%
\section{Conclusions and discussion}\label{sec: conclusions}
In this work, we investigate the effects of small localized perturbations on the spectrum (in)stability of total transmission modes (TTMs) of Tangherlini black holes. Previous pseudospectrum analyses have shown that TTMs of Tangherlini black holes are generally highly spectrally sensitive, with a notable exception being the purely imaginary mode~\cite{Zhou:2025xdo}. Motivated by these findings, we introduce explicit localized perturbations into the effective potential and directly track the resulting migration trajectories of the perturbed TTM spectra. This allows us to characterize the actual spectral response of TTMs to localized perturbations and to assess their spectrum stability from a perspective complementary to pseudospectrum analysis.

We model the localized perturbation using a P\"{o}schl-Teller bump with adjustable amplitude, width, and position, and formulate the TTM problem as a generalized eigenvalue problem. To address the numerical resolution difficulty caused by the narrowing of distant bumps in the compactified coordinate $\sigma$, we introduce a modified Chebyshev-Lobatto (CL) grid that concentrates collocation points around the bump, thereby effectively improving the numerical resolution of the localized perturbation.

For the $\text{TTM}_{\text{L}}$ with $d=14$, $\ell=2$, and $s=2$, the numerical results reveal two qualitatively distinct spectral responses. The purely imaginary $n=0$ mode remains confined to the imaginary axis, with its spectrum shift gradually decreasing as the bump is moved outward or the perturbation amplitude is reduced, indicating relatively strong spectrum stability [see Fig. \ref{fig:migration with alpha_11_2}]. In contrast, the complex-frequency $n=1$ and $n=2$ modes exhibit pronounced spectral sensitivity, with appreciable frequency migrations persisting even for perturbation amplitudes as small as $\epsilon=10^{-9}$ [see Fig. \ref{fig:migration with alpha_11_1}]. These spectral behaviors are consistent with the results reported in~\cite{Zhou:2025xdo}. Furthermore, the spectral migration depends strongly on the location of $V_\text{bump}$, but only weakly on its width. Although the spectrum sensitivity of the TTMs bears some resemblance to the environmental spectrum (in)stability of QNMs~\cite{Cheung:2021bol}, their migration trajectories do not exhibit the characteristic spiral structure observed for QNMs, indicating distinct responses of the two types of modes to distant localized perturbations.

Future work may extend this analysis to more general environmental perturbations and rotating black holes to test the generality of the TTM spectrum (in)stability. 

% Overall, our results reveal similarities between TTMs and QNMs in spectral sensitivity, while the purely imaginary TTM remains relatively stable under localized perturbations, consistent with previous pseudospectrum analyses. 

% This mode-dependent response may also affect the frequency-matching conditions for virtual absorption, providing new insights into black-hole scattering and absorption in realistic environments.

%======================================%
%<<<<<<<<<< Acknowledgement >>>>>>>>>>>%
%======================================%
\section*{Acknowledgement}
This work is supported by the National Natural Science Foundation of China under Grant No. 12505067. This work is also supported by the National Natural Science Foundation of China under Grant No. 12475067 and No. 12235019.

\appendix

\section{Eigenvalue method}\label{app: methods}
To solve TTMs problem namely Eq. (\ref{masterEq}) and boundary conditions (\ref{boundary_conditions}), following~\cite{Tuncer:2025dnp,Zhou:2025xdo}, we introduce  two new sets of coordinates $(t_{\pm},\sigma)$ for the first two dimensions $(t,r)$,
\begin{eqnarray}\label{EF_coor}
    t_{\pm}=\frac{1}{r_{\text{h}}}\Big[t\pm x(r)\Big]\, ,\qquad \sigma=\frac{r_{\text{h}}}{r}\, ,
\end{eqnarray}
where the $\pm$ corresponds to the $\text{TTM}_\text{R}$ and the $\text{TTM}_\text{L}$ case, respectively. Actually $t_{+}=(t+x)/{r_{\text{h}}}$ and $t_{-}=(t-x)/{r_\text{h}}$ are the dimensionless ingoing and outgoing Eddington-Finkelstein (EF) coordinates. From Eqs. (\ref{EF_coor}), after rescaling the field
\begin{eqnarray}\label{coord ts}
   {\psi}(\sigma)=\mathrm{e}^{\mp \mathrm{i}\omega_{\pm}x(\sigma)}\tilde{\psi}^{\pm}(\sigma)\, ,
\end{eqnarray}
$\Psi=\mathrm{e}^{-\mathrm{i}\omega t}\psi$ then becomes
\begin{eqnarray}\label{coord ts1}
   \Psi(t_{\pm},\sigma)=\mathrm{e}^{-\mathrm{i}\omega_{\pm}r_{\text{h}}t_{\pm}}\tilde{\psi}^{\pm}(\sigma)\, .
\end{eqnarray}
Note that $\omega_{+}$ stands for the $\text{TTM}_\text{R}$ and $\omega_{-}$ stands for the $\text{TTM}_\text{L}$. Using Eqs. (\ref{EF_coor}) and Eq. (\ref{coord ts1}), Eq. (\ref{masterEqTime}) then becomes
\begin{eqnarray}\label{Compactified Eq}
    L_1\tilde{\psi}^{\pm}=\mp\mathrm{i}\omega_{\pm} r_{\text{h}}L_2\tilde{\psi}^{\pm}\, ,
\end{eqnarray}
where
\begin{eqnarray}\label{TTMs Eq}
    L_1=\frac{\mathrm{d}}{\mathrm{d}\sigma}\Big(p(\sigma)\frac{\mathrm{d}}{\mathrm{d}\sigma}\Big)-\frac{r_\text{h}^2 V(\sigma)}{p(\sigma)}\, ,\qquad L_2=2\frac{\mathrm{d}}{\mathrm{d}\sigma}\, ,\quad \text{with} \quad p(\sigma)=\sigma^2 f(r(\sigma))\, .
\end{eqnarray}
Here, the potential $V(\sigma)$ in $L_1$ should be instead of the modified potential $V_\epsilon(\sigma)=V(\sigma)+V_{\text{bump}}(\sigma)$. Because the function $p(\sigma)$ vanishes at both boundaries, TTMs are formulated in terms of the regular solutions $\tilde{\psi}$ of Eq. (\ref{Compactified Eq}). Eq. (\ref{Compactified Eq}) is solved numerically by discretizing the differential operators $L_{1}$ and $L_{2}$ into matrices using a Chebyshev-Lobatto grid associated with resolution $N$~\cite{Jaramillo:2020tuu,Jansen:2017oag,Yang:2025hqk}. Owing to spherical symmetry, the TTM spectra are symmetric about the imaginary axis, while $\text{TTM}_\text{L}$ and $\text{TTM}_\text{R}$ are related by reflection about the real axis. Therefore, we only consider $\text{TTM}_\text{L}$ modes.

\section{Modified Chebyshev-Lobatto grid}\label{app: modified CL}
From Eq. (\ref{s perturbation}), we can see that small bump of the PT form is imposed in tortoise coordinate. After compacting the coordinate, the width of the bump in $\sigma$ coordinate will become narrower and narrower as $a\to+\infty$. Using the usual Chebyshev-Lobatto (CL) grid can not describe such bump unless the resolution is very high. To alleviate the above issue, we provide a reliable method in this appendix. First, we begin with the standard CL grid, i.e., 
\begin{eqnarray}\label{standard_CL_grid}
   x_j=\cos\Big(\frac{j\pi}{N}\Big)\, ,\qquad j=0,1,\cdots,N-1,N\, ,
\end{eqnarray}
where $x_j\in[-1,1]$. Since the standard CL grid is naturally clustered near the interval endpoints $x=\pm1$, they may provide insufficient resolution for a narrow structure located in the interior of the computational domain. To improve the numerical resolution of the localized bump $V_{\text{bump}}$, we introduce a modified CL grid in the followings. A transformation from $[-1,1]$ to $[-1,1]$ given by
\begin{eqnarray}\label{q_x}
   q(x)=\frac{\sinh(\beta x)}{\sinh\beta}\, ,\quad x\in[-1,1]\, ,
\end{eqnarray}
where $\beta>0$ controls the strength of the redistribution is introduced, and $\beta=0$ can be regarded as the original CL grid. Similar manipulations can also be found in~\cite{Siqueira:2025lww}. This transformation preserves the endpoints, namely $q(\pm1)=\pm1$, and one also has $q(0)=0$. One can increase $\beta$ to increase the density of grid points at $0$. We visualize the impact of different $\beta$ on grid distribution within the interval $[-1,1]$ in Fig. \ref{Gridposition}.

\begin{figure}[htbp]
    \centering
    \includegraphics[width=0.7\textwidth]{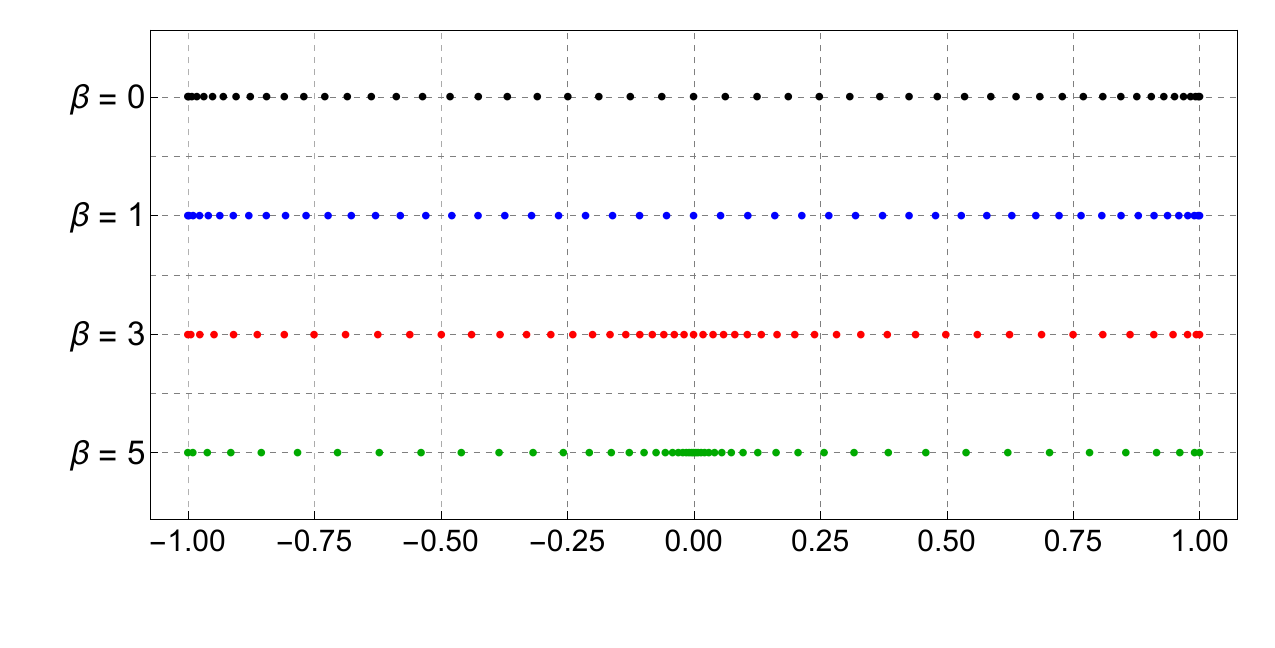}
    \caption{Distributions of the original CL grid $(\beta=0)$ and the modified CL grids for $\beta=1$, $3$ and $5$, in which the resolution is $N=50$. We can see that as $\beta$ increases, the grid points gradually converge towards $x=0$.}
    \label{Gridposition}
\end{figure}

In terms of the coordinate $\sigma$, the center of $V_{\text{bump}}$ is denoted by $\sigma_c$. In order to place the high-resolution region at the center $\sigma_c$, we further introduce the following M\"{o}bius transformation
\begin{eqnarray}\label{Mobius_transformation}
   \sigma(q)=\frac{\sigma_c(1+q)}{1+(2\sigma_c-1)q}\, ,
\end{eqnarray}
which maps $q\in[-1,1]$ onto the physical domain $\sigma\in[0,1]$. In particular, the transformation satisfies $\sigma(-1)=0$, $\sigma(0)=\sigma_c$ and $\sigma(1)=1$. Moreover, one has
\begin{eqnarray}
    \frac{\rm{d}\sigma}{\mathrm{d}q} = \frac{2\sigma_c(1-\sigma_c)} {\big[1+(2\sigma_c-1)q\big]^2}>0\, , \qquad q\in[-1,1]\, ,
\end{eqnarray}
for $0<\sigma_c<1$. Hence, $\sigma(q)$ is smooth and strictly monotonically increasing throughout the computational domain and therefore defines a one-to-one bijective mapping from $[-1,1]$ to $[0,1]$. In particular, the clustering center $q=0$ [see Eq. (\ref{q_x})] is mapped precisely to the center of the localized bump, $\sigma=\sigma_c$. Accordingly, the grid points clustered around $q=0$ are mapped to the vicinity of $\sigma=\sigma_c$.

So far, we have obtained a new grid point distribution characterized by denser grid points near the bump. What needs to be done now is to provide the differential matrix under this grid point distribution. The differential matrix $\mathbf{D}_x$ is constructed with respect to the original CL grid (\ref{standard_CL_grid}). New differential matrices in the physical coordinate $\sigma$ are able to be obtained by the chain rule. From Eq. (\ref{q_x}) and Eq. (\ref{Mobius_transformation}), one gets the function $\sigma(x)$ with $\sigma(x)=\sigma(q(x))$. According to
\begin{eqnarray}
    \frac{\mathrm{d}}{\mathrm{d}\sigma} = \frac{1}{\sigma^{\prime}(x)} \frac{\mathrm{d}}{\mathrm{d}x}\, ,\quad \text{and} \quad \frac{\mathrm{d}^2}{\mathrm{d}\sigma^2} = \frac{1}{\big(\sigma^{\prime}(x)\big)^2}\frac{\mathrm{d}^2}{\mathrm{d}x^2} - \frac{\sigma^{\prime\prime}(x)}{\big(\sigma^{\prime}(x)\big)^3}\frac{\mathrm{d}}{\mathrm{d}x}\, ,
\end{eqnarray}
new differential matrices are given by
\begin{eqnarray}\label{new_differential_matrices}
    \mathbf{D}_{\sigma}=\text{diag}\Big(\frac{1}{\sigma^{\prime}(x)}\Big)\cdot\mathbf{D}_x\, ,\quad \text{and} \quad \mathbf{D}_{\sigma\sigma}=\text{diag}\Big(\frac{1}{\sigma^{\prime}(x)^2}\Big)\cdot\mathbf{D}_{x}^2-\text{diag}\Big(\frac{\sigma^{\prime\prime}(x)}{\sigma^{\prime}(x)^3}\Big)\cdot\mathbf{D}_x\, .
\end{eqnarray}
The modified grid therefore allocates a larger fraction of the collocation points to the neighborhood of the localized bump, improving its numerical resolution without increasing the total number of grid points.
\bibliography{reference}
\bibliographystyle{apsrev4-1}

\end{document}